\documentclass[graybox, envcountchap]{svmult}

\usepackage{mathptmx}        
\usepackage{amsmath}
\usepackage{amssymb}
\usepackage{color}
\usepackage{helvet}          
\usepackage{courier}         
\usepackage{dirtree}

\usepackage{makeidx}        
\usepackage{graphicx}        
\usepackage{subfig}

\usepackage{multicol}        
\usepackage[bottom]{footmisc}

\usepackage{hyperref}        
\hypersetup{colorlinks=true,urlcolor=blue}

\usepackage[misc]{ifsym}

\makeindex             

\begin{document}

\title{CMB Anomalies and the Hubble Tension}

\author{William Giar\`e}
\institute{William Giar\`e (\Letter) \at School of Mathematics and Statistics, University of Sheffield, Hounsfield Road, Sheffield S3 7RH, United Kingdom\\ \email{w.giare@sheffield.ac.uk}}
%
%
\maketitle

\abstract{The standard $\Lambda$CDM model of cosmology is largely successful in describing many observations, including precise measurements of the Cosmic Microwave Background (CMB) radiation. However, some intriguing anomalies remain currently unexplained within this theoretical framework. Such discrepancies can be broadly categorized into two groups: those involving CMB-independent probes and those within different CMB experiments. Examples of the former category include the $H_0$-tension between the value of the present-day expansion rate of the Universe inferred by CMB observations and local distance ladder measurements. The latter category involves anomalies between the values of cosmological parameters obtained by different CMB experiments, as well as their consistency with the predictions of $\Lambda$CDM. In this chapter, we primarily focus on this second category and study the agreement among the most recent CMB measurements to acquire a better understanding of the limitations and uncertainties underlying both the current data and the cosmological model. Finally, we discuss the implications for the $H_0$-tension.}

\section{CMB Anomalies: A Brief Multi-Experiment Overview}
\label{sec.anomalies}

The most recent measurements of the Cosmic Microwave Background temperature and polarization anisotropies released by the Planck Collaboration~\cite{Planck:2018nkj} have achieved sub-percent accuracy in extracting most of the cosmological parameters~\cite{Planck:2018vyg}. While this is undoubtedly a blessing, it also represents a challenge: as precision improves, any deviations or anomalies may become more statistically significant and potentially point to tensions in our current understanding of the Universe~\cite{Abdalla:2022yfr,DiValentino:2022fjm,Perivolaropoulos:2021jda}.

One notable example is the higher lensing amplitude at about $2.8\sigma$ observed in the Planck data~\cite{DiValentino:2019dzu,DiValentino:2020hov,Motloch:2018pjy}. Since more lensing is expected with more Cold Dark Matter (CDM), the lensing anomaly immediately recast a preference for a closed Universe, which in turn helps to explain other large-scale anomalies in the data, such as the deficit of amplitude in the quadrupole and octupole modes. Consequently, the final Planck indication for a closed Universe becomes very significant, reaching the level of $3.4\sigma$~\cite{DiValentino:2019qzk,Handley:2019tkm}.

Excitingly, the lensing and curvature anomalies, as well as any other observed discrepancies~\cite{Fosalba:2020gls,Schwarz:2015cma,Yeung:2022smn,Haslbauer:2020xaa}, may indicate the presence of new physics beyond the standard $\Lambda$CDM model of cosmology. However, it is also entirely plausible that anomalies simply reflect observational systematic errors and/or statistical fluctuations in the data. One effective way to discriminate between the two possibilities is to compare the results of multiple CMB measurements: any persistent deviations across independent probes may strongly hint at the presence of new physics. Fortunately, this is now possible thanks to the observations of the CMB angular power spectra recently released by the Atacama Cosmology Telescope (ACT)~\cite{ACT:2020gnv,ACT:2020frw} and the South Pole Telescope (SPT)~\cite{SPT-3G:2022hvq,SPT-3G:2014dbx,SPT-3G:2021eoc}. While these experiments may not be as constraining as the Planck satellite, they are still able to produce cosmological bounds that are accurate enough for independent tests of the Planck results~\cite{ACT:2020gnv,SPT:2017sjt,SPT-3G:2021wgf,SPT-3G:2022hvq}. 

Interestingly, both the ACT and SPT data have provided full support for the inflationary prediction of a spatially flat Universe and for a lensing amplitude consistent with $\Lambda$CDM~\cite{ACT:2020gnv,SPT-3G:2021wgf,SPT-3G:2022hvq}. This suggests that both the curvature and lensing anomalies may be explained in terms of statistical fluctuations or systematic errors. However, the same datasets have revealed other relevant deviations from the standard cosmological model. For instance, the Data Release 4 of the Atacama Cosmology Telescope shows preference for a unitary spectral index of primordial perturbations ($n_s=1.009 \pm 0.015$)~\cite{ACT:2020gnv,Giare:2022rvg}, in agreement with the phenomenological scale-invariant spectrum proposed by Harrison, Zel’dovich, and Peebles~\cite{Harrison:1969fb,Peebles:1970ag,Zeldovich:1972zz}, and in tension with Planck, as well as with the predictions of the most typical single-field inflationary models~\cite{Giare:2023wzl}, at more than $99.3\%$~Confidence Level (CL). Other mild yet relevant anomalies in the ACT data concern the value of the effective number of relativistic degrees of freedom in the early Universe ($N_{\rm{eff}}=2.46\pm0.26$)~\cite{ACT:2020gnv, DiValentino:2022oon}, in disagreement at more than 2$\sigma$ with the expected value of $N_{\rm{eff}}=3.04$ predicted by the Standard Model of particle physics for three active neutrinos~\cite{Bennett:2020zkv}; a $\sim 2\sigma$ preference for a positive running of the scalar spectral index of inflationary perturbations ($\alpha_{\rm s}\equiv dn_s/d\log k= 0.058\pm 0.028$)~\cite{ACT:2020gnv,Forconi:2021que,Giare:2022rvg};
and a $\sim 3\sigma$ indication in favor of Early Dark Energy~\cite{Hill:2021yec,Poulin:2023lkg}.

In turn, these deviations are not supported by Planck, which instead shows $\sim 8\sigma$ evidence for $n_{\rm s}\ne 1$, a good agreement with the predictions of the standard model for $N_{\rm eff}$, and no evidence for Early Dark Energy~\cite{Planck:2018vyg,Hill:2021yec,Poulin:2023lkg}. Therefore, also in this case, the lack of cross-validation between independent CMB measurements may suggest statistical fluctuations and/or as of yet unknown systematic errors. 

Regardless of whether these anomalies arise from observational systematic effects, statistical fluctuations, or new physics, if we aim to achieve precision cosmology and test fundamental physics, it is important to acquire a better comprehension of the collective agreement between independent experiments and to have a clear understanding of the limitations and uncertainties underlying both current data and the cosmological model. This becomes a crucial need in relation to the $H_0$-tension~\cite{Bernal:2016gxb,Verde:2019ivm} as many proposed solutions\footnote{See, \textit{e.g.}, Refs.~\cite{Abdalla:2022yfr,DiValentino:2021izs,Jedamzik:2020zmd,Kamionkowski:2022pkx,Knox:2019rjx,Poulin:2023lkg,Krishnan:2020vaf,Colgain:2021beg} for recent reviews and discussion.} call for a new paradigm shift in cosmology, while relying almost entirely on the resilience of such observations.

In this chapter, we address this problem and discuss the global consistency of independent CMB experiments. In \autoref{sec.GlobalTension}, we investigate potential reasons for discrepancies by exploring both the limitations inherent in current data (\autoref{sec.data}) and the uncertainties of the standard cosmological model (\autoref{sec.model}). We eventually argue that anomalies seem to indicate a potential mismatch between the early and late Universe (\autoref{sec.howmany}). In \autoref{sec.H0}, we discuss the implications for the $H_0$-tension and its most common solutions.


\section{Global Consistency of CMB experiments}
\label{sec.GlobalTension}

Based on the overview provided in the previous section, it is evident that what makes CMB anomalies difficult to interpret \textit{individually} is that different experiments often point in discordant directions, and none of the most relevant deviations can be cross-validated through independent probes. For this reason, accurate statistical methods have been developed to quantify the \textit{global} consistency between independent experiments under a given model of cosmology, for instance by starting from the constraints on the parameter space of the model itself~\cite{Calderon:2023obf,Charnock:2017vcd,DiValentino:2022rdg,Handley:2019wlz,LaPosta:2022llv,Lin:2019zdn}.

Assuming a standard $\Lambda$CDM cosmology, the data release 4 of the Atacama Cosmology Telescope can be evaluated in mild-to-moderate tension with Planck and the South Pole Telescope at a gaussian equivalent level of 2.6$\sigma$ and 2.4$\sigma$, respectively~\cite{DiValentino:2022rdg,Handley:2020hdp}. Depending on one's level of tolerance, these controversial tensions ranging between $2-3\sigma$ may be taken more or less seriously. If our objective is to accomplish "precision cosmology" and effectively assess fundamental physics, it is reasonable to conclude that they warrant further investigations. However, it remains crucial to understand the causes of these differences and shed light on their potential implications for the numerous proposed resolutions of the $H_0$-tension that are emerging from these observations.

\subsection{The limitations of current Data}
\label{sec.data}

Assuming a baseline $\Lambda$CDM cosmology, the main source of tension between ACT and Planck arises from the measurements of the scalar spectral index, $n_{\rm s}$, with Planck reporting a value of $n_{\rm s}=0.9649\pm0.0044$~\cite{Planck:2018vyg} and ACT producing a value of $n_{\rm s}=1.008\pm0.015$~\cite{ACT:2020gnv}; in tension at approximately $2.7\sigma$. Moreover, the discrepancy between the two experiments extends to the measurement of the baryon energy density with Planck and ACT giving respectively $\Omega_b h^2=0.02236\pm0.00015$~\cite{Planck:2018vyg} and $\Omega_b h^2= 0.02153 \pm 0.00030$~\cite{ACT:2020gnv}; in tension at about $2.5\sigma$. 

If we believe these differences to emerge from limitations in the data, then a logical step is to identify which (missing) part of the dataset is responsible for the discrepancy and eventually discard (include) it in the cosmological parameter inference analyses. Several possible explanations have been attempted to account for the disagreement.

\begin{itemize}

\item \textbf{Temperature data}: One suggested explanation could be attributed to the limited amount of ACT data at low multipoles in the TT Spectrum where ACT has a minimum sensitivity in multipole of $\ell=600$~\cite{ACT:2020gnv}. In absence of data around the first two acoustic peaks, a strong degeneracy exists between $\Omega_b h^2$ and $n_{\rm s}$ as a lower value of the former can be mimicked by a larger value of the latter, tilting the spectrum in the opposite direction~\cite{ACT:2020gnv}. Typically, the lack of information is filled by using measurements from Planck or a third independent experiment such as WMAP~\cite{WMAP:2012fli,Hinshaw:2012aka} in the multipole range $\ell \lesssim 600$. While this helps in diluting the tension, it is worth noting that there is no a-priori reason to combine ACT with low-$\ell$ data from another experiment: despite the degeneracy, accurate data at the scales probed by ACT are enough to achieve precise constraints on both $n_{\rm s}$ and $\Omega_{\rm b}h^2$. For instance, considering the Planck data only in the multipole range in common with ACT, the final results still indicate $n_{\rm s}\ne 1$ at more than $4\sigma$~\cite{Giare:2022rvg}. This cast doubts on the explanation of the mismatch in terms of lack of information on the first acoustic peaks as one would expect a reasonable agreement between two experiments measuring an overlapping range of multipoles. \\

\item \textbf{Polarization data}: Another possibility is that the limited amount of ACT low multipoles polarization data (which cover only multipoles $\ell \gtrsim 350$ in TE and EE) is responsible for the discrepancy. However, incorporating low-multipole polarization data from the Planck measurements of E-modes at $2\le \ell \le 30$ fails to explain these anomalies~\cite{Giare:2022rvg}. Nonetheless, it was also suggested that an overall TE calibration could resolve the mismatch~\cite{ACT:2020gnv}. Neglecting any information from ACT polarization measurements (TE EE) and combining ACT temperature anisotropies (TT) with SPT polarization data (TE EE) the tension is in fact reduced below $2\sigma$, restoring the agreement between the experiments for $\Omega_b h^2$~\cite{Giare:2022rvg}. However, the same combination of ACT and SPT data still favors a value of $n_{\rm s}$ around unity~\cite{Giare:2022rvg}, indicating that ACT polarization may only partially contribute to the tension, without fully reconciling it.\\

\item \textbf{Non-CMB data}: Yet another possible avenue to increase the data constraining power is to break the geometrical degeneracy among cosmological parameters by exploiting CMB-independent data, such as astrophysical measurements of Baryon Acoustic Oscillation (BAO) and Redshift Space Distortion (RSD)~\cite{BOSS:2012dmf,Dawson:2015wdb}, as well as galaxy clustering and cosmic shear observations~\cite{DES:2017myr,DES:2021wwk}. While astrophysical data have been found to alleviate several tensions and restore the parameter values to those predicted by $\Lambda$CDM, this is not the case. Including them does not shift the results significantly but leads to tighter errors, ultimately increasing the tension~\cite{ACT:2020gnv,Giare:2022rvg}.
\end{itemize}

\begin{figure*}[t]
\includegraphics[width=\textwidth]{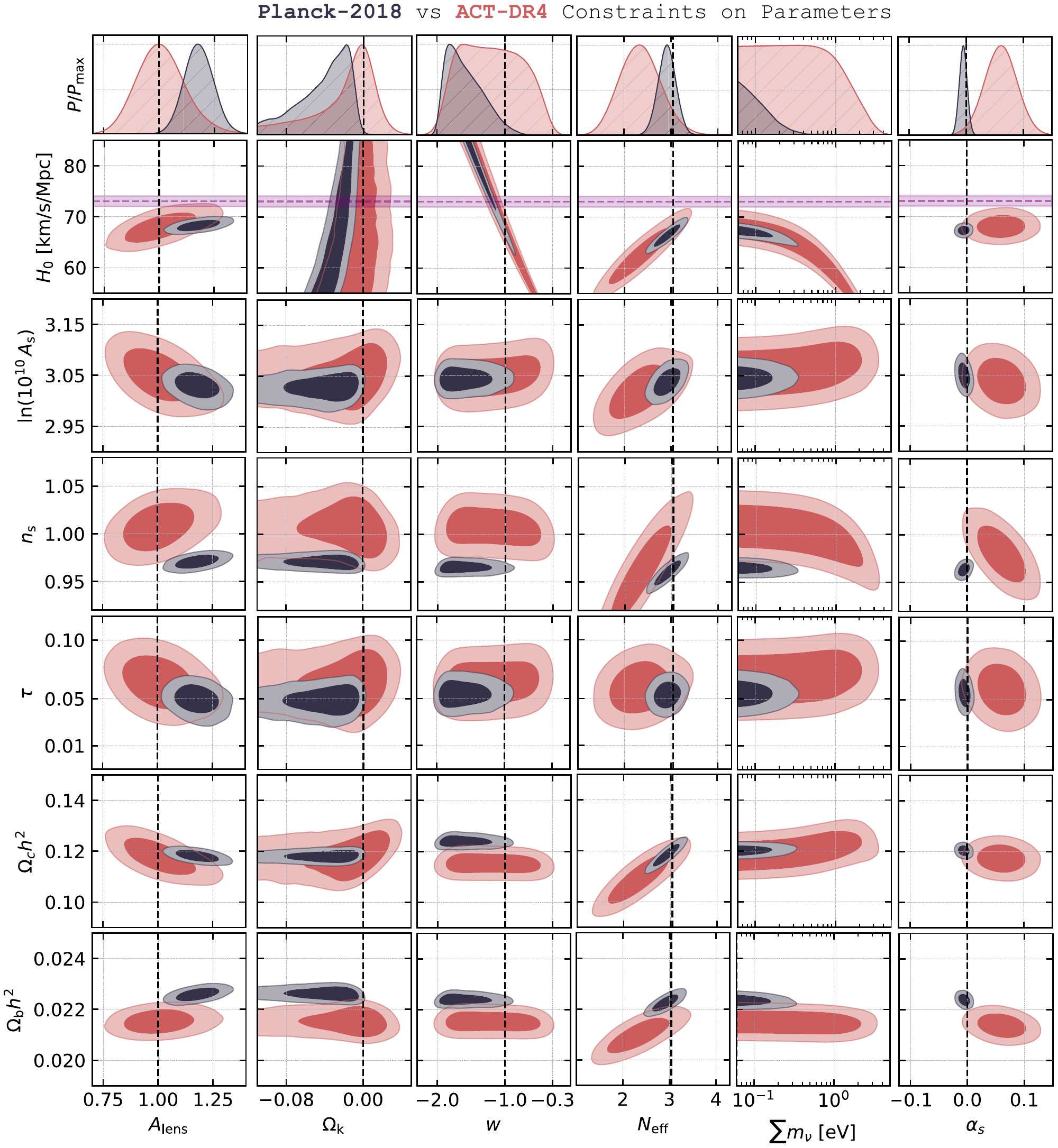}
\caption{\small Constraints on cosmological parameters in various minimal extensions to $\Lambda$CDM derived from the analysis of the Planck 2018~\cite{Planck:2018vyg} (grey) and ACT-DR4~\cite{ACT:2020gnv} (red) temperature and polarization data. The vertical dashed black lines indicate the reference values of parameters in $\Lambda$CDM. The horizontal magenta band in the 2D-plots involving the Hubble constant shows the direct measurements obtained by the SH0ES Collaboration ($H_0=73\pm1$ km/s/Mpc)~\cite{Riess:2021jrx}. }
\label{fig:1} 
\end{figure*}

\subsection{The unknowns of the Cosmological Model}
\label{sec.model}

Although observational systematic errors and/or statistical fluctuations in the data may provide a simple explanation for the discrepancies between different CMB experiments, we cannot discount the possibility that part of these anomalies are rooted in the limitations of the standard cosmological model to accurately explain precise observations at very different angular scales. 

From a theoretical standpoint, many ingredients of $\Lambda$CDM lack solid understanding in fundamental physics. The nature of Dark Matter and Dark Energy remains unknown in the context of the Standard Model of fundamental interactions as well as the physical origin of the inflationary epoch is also unclear. To account for all these unknowns, the $\Lambda$CDM model adopts a principle of simplicity and incorporates these components into a streamlined theoretical framework consisting of just six free parameters. However, the value of cosmological parameters inferred from CMB measurements clearly depends on the global parameterization adopted for the background cosmology and - ultimately - on the model’s assumptions. For this reason, a possibility usually explored when finding anomalies in the cosmological parameter values, is to extend the baseline model and study how the overall tension changes. 

\autoref{fig:1} provides a comprehensive summary of the constraints on cosmological parameters derived from Planck and ACT in various minimal extensions to $\Lambda$CDM, including those involving the anomalous parameters discussed so far. Directing our attention to the enigmatic foundations of the standard cosmological model, a careful inspection on the figure can yield valuable insights into the uncertainties underlying their theoretical parameterization and potentially reveal any missing ingredients in the theory able to account for the observed discrepancies.

\begin{itemize}
    \item \textbf{Dark Energy:}  Within the standard cosmological model, the late-time accelerated expansion of the Universe is attributed to a cosmological constant term in the Einstein equations, resulting in a constant energy-density component with a negative equation of state, $w = -1$. Relaxing this assumption we can use the Planck and ACT data to directly constrain $w$ and test its consistency with the cosmological constant value. Interestingly, Planck suggests a phantom equation of state ($w<-1$) at about $2\sigma$, while ACT is consistent with $w = -1$, albeit with large uncertainties that cannot exclude phantom or quintessential ($w>-1$) behaviors, see \autoref{fig:1}. In any case, late-time modifications can hardly reconcile the difference between the two experiments, as it mainly arises from a mismatch of the early Universe.
    \\

    \item \textbf{Dark Matter and Neutrinos:} As the tension between Planck and ACT arises primary from a mismatch in the values of early-time parameters, interesting extensions are those associated with the early Universe content, such as neutrinos, Dark Matter and Dark radiation. For instance, relaxing the value of total neutrino mass (fixed at $\sum m_{\nu}\sim 0.06$ eV within $\Lambda$CDM) produces a shift in the value of $n_s$ measured by ACT towards the Planck result. However, such a shift can be associated with the ACT's peculiar preference for higher values of neutrino mass, which contradicts the tight constraints placed by the Planck data.
    A second an interesting avenue involves the radiation energy-density in the early Universe. In the case of three active neutrinos, the radiation energy density can be calculated quite accurately within the standard model of particle physics and parameterized in terms of the effective number of relativistic degrees of freedom, $N_{\rm eff}=3.044$. By relaxing this theoretical constraint and leaving $N_{\rm eff}$ as a free parameter to be inferred by data, we can use the ACT and Planck observations to directly verify the consistency with the Standard Model's predictions as well as to test any theoretical extensions involving additional (thermal) degrees of freedom in the early Universe. From \autoref{fig:1} we see that Planck is in good agreement with the Standard Model predictions, while the ACT's result ($N_{\rm{eff}}=2.46\pm0.26$) is anomalous and in disagreement with the expected value for three active neutrinos by about $2.5\sigma$. Interestingly, when the energy density in the early Universe is significantly lower than the expected value, the disagreement between the two experiments gets diluted by the simultaneous effect of widening the uncertainties and shifting the values of $n_s$ towards that measured by Planck. The interpretation of this result highly depends on our attitude regarding the reasons for the disagreement. If we attribute the observed discrepancy between ACT and Planck to new physics, this may suggest a potential role for the neutrino sector (and generally for new physics at early times) in accounting for at least part of this difference. On the other hand, if we rely primarily on the strength of the theoretical predictions, this may indicate that observational systematic effects are introducing biases into the determination of $N_{\rm eff}$, raising concern on the results obtained when considering new physics at early times. 
    \\
    
    \item \textbf{Inflation:} One of the most common results from single-field models of inflation is the production of nearly scale-invariant perturbations that can be well described by a power-law form of the adiabatic scalar components. In the simplest scenarios, any residual scale dependence is parameterized solely in terms of the scalar spectral index $n_{\rm s}$, where $n_s=1$ corresponds to exact scale-invariance. However, additional parameterizations that incorporate higher-order terms in the power-law expansion should also be considered. In this regard, another important clue that can be inferred from \autoref{fig:1} is the tendency for the value of $n_s$ predicted by ACT to decrease significantly when extending the inflationary sector of the theory by accounting also for the possibility of a running in the scalar spectral index $\alpha_{\rm s}= d n_{\rm s} / d\log k$. In this case, the difference on the spectral index $n_{\rm s}$ is solved, but the disagreement maps into a strong controversy in the values of $\alpha_{\rm s}$: while Planck prefers a slightly negative running ($\alpha_{\rm s}= -0.0119\pm 0.0079$), ACT favors a positive running ($\alpha_{\rm s}= 0.058\pm 0.028$).\\

\end{itemize}

\subsection{How many early-late Universe mismatches are there?}
\label{sec.howmany}

To draw some general conclusions from the precedent analysis and gain a fresh perspective on the CMB anomalies, we want to argue that the global tension emerging between ACT and Planck always maps into mismatches between the early and late Universe. 

Considering all the minimal extensions in \autoref{fig:1}, we can notice that the Planck anomalies \textit{always} involve parameters associated with the local Universe such as the lensing amplitude, the spacetime geometry, and the dark energy equation of state. Instead, the ACT anomalies \textit{always} involve parameters related to the early Universe such as the baryon energy density, the spectral index, its running, and a free-to-vary neutrino sector. Therefore, the picture emerging is the following one: Planck shows mild-to-moderate deviations from $\Lambda$CDM when extending the late time sector of the theory, remaining consistent for extensions involving modifications at early-times.  In contrast, the Atacama Cosmology Telescope agrees pretty well with $\Lambda$CDM at late-times, but shows anomalies in the early Universe parameters. However, considering the large experimental uncertainties obtained when extending the late-time sector of the theory, the difference between the two probes remains mostly driven by a mismatch in the early Universe.


\section{Implications for the $H_0$-Tension}
\label{sec.H0}

Any determination of the Hubble constant inferred from observations of the cosmic microwave background necessarily depends on the cosmological model. For this reason, several alternative models have been proposed to address the widely known tension between the \textit{direct} value of $H_0$ measured by the SH0ES collaboration using luminosity distances of Type Ia supernovae~\cite{Riess:2021jrx} ($H_0=73\pm1$ Km/s/Mpc) and the \textit{indirect} value inferred by the Planck satellite~\cite{Planck:2018vyg} ($H_0=67.4\pm0.5$ Km/s/Mpc).

If we intend to explain the mismatch in terms of new physics (and clarify the potential role of the CMB anomalies in this matter), it is important to understand what aspects of the theory are involved in the game when inferring $H_0$ from observations of the CMB. It is well known that measuring the positions of the acoustic peaks in the CMB angular power spectrum, we can determine the angular size of the sound horizon ($\theta_s$) which is the cosmological parameter most accurately measured by Planck~\cite{Planck:2018vyg}. On the other hand, we can fix the baryon density $\Omega_b h^2$ from the relative heights of even and odd peaks, while we can obtain the cold dark matter density $\Omega_c h^2$ from the absolute heights of acoustic peaks. Once these parameters are known, we can pick a model of the early Universe (i.e., pre-recombination) and calculate the physical dimension of sound horizon ($r_s$). Given $r_s$ and $\theta_s$, we can easily infer the angular diameter distance from the CMB, $D_A(z_{ls}) = r_s/\theta_s$ (with $z_{ls}\simeq 1100$). Finally, by assuming a cosmological model for the subsequent late-time expansion (i.e., post-recombination), we eventually infer the value of $H_0$ from $D_A(z_{ls})$. 

Therefore, any theoretical attempts aimed at solving the $H_0$-tension should fall into one or both these categories: 

\begin{figure*}[t]
\includegraphics[width=\textwidth]{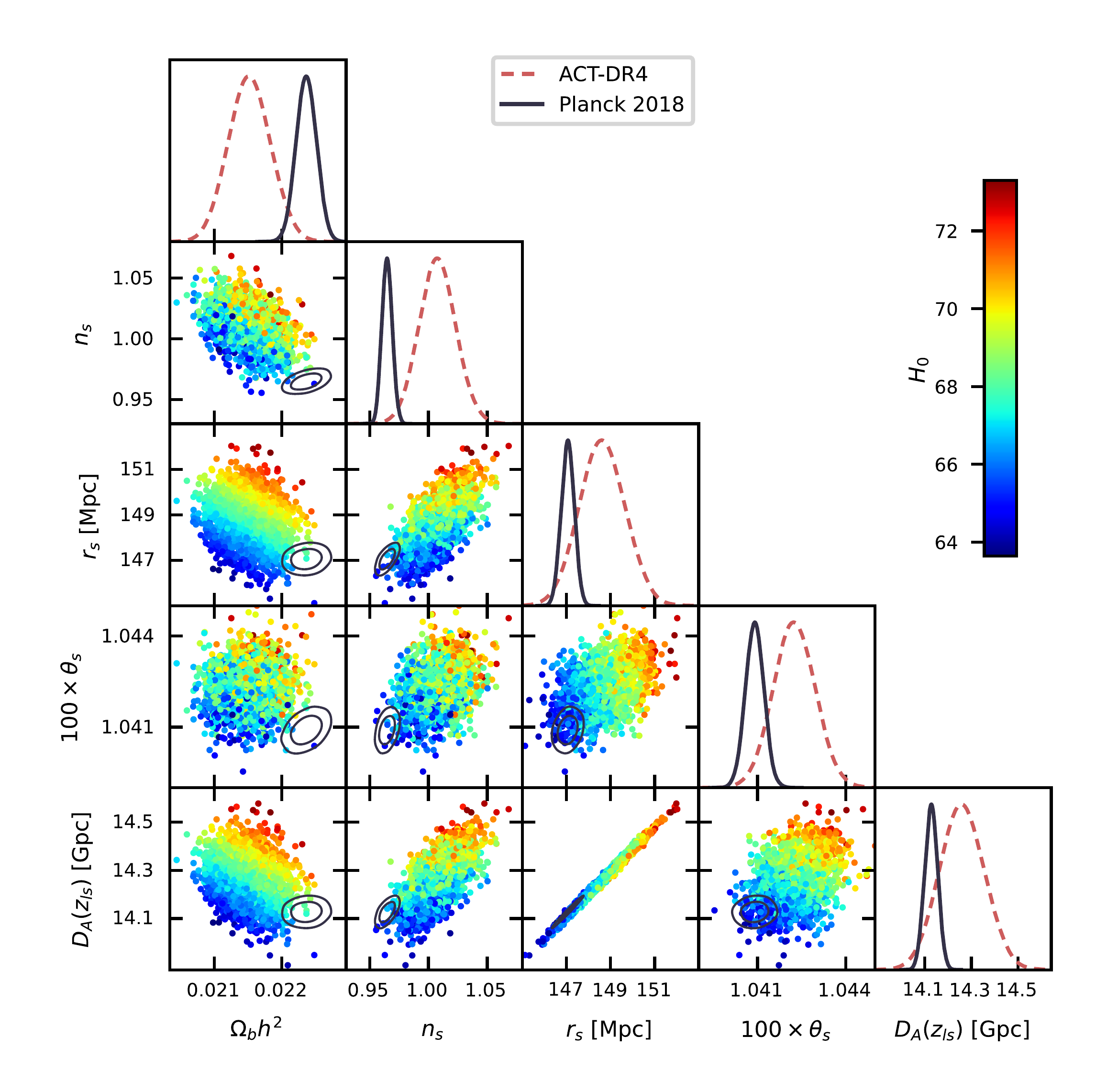}
\caption{\small Constraints on the early Universe parameters and their relation with $H_0$ (in the color-bar) as obtained by the ACT-DR4 data release~\cite{ACT:2020gnv}. The black empty contours define the region of the parameter space preferred by the Planck 2018 data~\cite{Planck:2018vyg}.}
\label{fig:2} 
\end{figure*}

\begin{itemize}
\item \textbf{Late time solutions:} The first possibility is to modify the late time expansion history and change the value of the Hubble constant inferred from the angular diameter distance $D_A(z_{ls})$. This is typically done by introducing new late-time physics and/or extending the dark sector of the theory. Based on \autoref{fig:1}, one might expect these solutions to be preferred by data, given the significant room left by the CMB observations for new physics at late-times. However, they should be approached with some caution as late-time modifications often leads to a strong geometric degeneracy among parameters which needs to be broken in order to genuinely test new physics. In practice, this can be done by incorporating local Universe observations, such as baryon acoustic oscillations~\cite{BOSS:2012dmf,Dawson:2015wdb} and/or supernovae measurements~\cite{Brout:2022vxf,Scolnic:2021amr}. They often clean away any late-time anomalies and restore the parameter values to $\Lambda$CDM. Therefore, contrary to what one might expect from CMB data, there is very little room to accommodate new physics at late-times. However, it is worth considering that the inclusion of certain local probes becomes a subject of debate when testing theoretical models that significantly deviate from $\Lambda$CDM. For instance, this objection is sometimes raised against BAO measurements, which assume a $\Lambda$CDM-like cosmology as an ansatz from the outset, but are in tension with CMB in many extended theoretical frameworks~\cite{DiValentino:2019qzk,Handley:2019tkm}. Therefore, 
taking a conservative approach, we cannot discount the possibility that (part of) the $H_0$-tension may be alleviated by late-time solutions. In any case, it is unlikely that the tension between ACT and Planck will have a significant impact on these solutions since these experiments primarily disagree at early times, as seen in \autoref{sec.GlobalTension}.
\\

\item \textbf{Early time solutions:} The second possibility is to consider new physics in early Universe to change the physical size of the sound horizon $r_s$. This is typically done by incorporating new light degrees of freedom before recombination, decreasing $r_s$ and raising $H_0$. However, these solutions come with their own caveats. For instance, many indications of this kind of new early-time physics arise when combining multiple CMB measurements at different angular scales (such as Planck and ACT), without finding clear cross-validation when these experiments are considered separately~\cite{Hill:2021yec,Poulin:2023lkg}. Therefore, this may cast doubt on whether these indications genuinely reflect new physics or if they emerge from the controversial tension between these probes, as also seen in \autoref{sec.model} for many simpler minimal early-Universe extensions. To explore this issue further, we remain agnostic and refer to \autoref{fig:2}. Assuming a background $\Lambda$CDM cosmology, the figure depicts the constraints on the early Universe parameters of interest for early-time solutions, showing a consistent shift between ACT and Planck. Notably, ACT allows for greater flexibility in accommodating higher values of the sound horizon, while Planck peaks where ACT prefers very low values of $H_0$. This means that increasing $H_0$ requires moving towards the region of the parameter space where the disagreement between the two probes becomes more significant. In addition, the spectral index and the Hubble constant are all positively correlated: increasing $H_0$ will naturally push $n_s$ towards higher values. This is why many potential solutions to the $H_0$ tension suggest a prominent role of $n_s$, often pointing towards values of $n_s\sim 1$ as measured by ACT, bringing the tension around this parameter back on the table, see, \textit{e.g.}, Refs.~\cite{Jiang:2022uyg,Jiang:2022qlj,Ye:2022efx,Jiang:2023bsz}. However, one might argue that relaxing the early Universe sector of the theory will change the correlation among parameters. As seen in \autoref{sec.model}, considering a free-to-vary $N_{\rm eff}$ can partially accommodate the disagreement between ACT and Planck. While this is true, the results in \autoref{fig:1} show a significant shift of $N_{\rm eff}$ towards smaller values, which is in the opposite direction of what we need to solve the $H_0$ tension. This also raises concerns about some proposed solutions whose underlying philosophy consists of incorporating \textit{large} additional degrees of freedom as they appear to be disfavored in the first place.

\end{itemize}

\bibliographystyle{acm}
\bibliography{biblio}
\end{document}